\documentclass[11pt,a4paper]{article}
\pdfoutput=1
\usepackage{empheq}
\usepackage{cases}
\usepackage{a4wide}
\usepackage{amsfonts}
\usepackage{amssymb}
\usepackage{amsmath}
\usepackage{graphicx}
\usepackage{booktabs}
\usepackage{array}
\usepackage{color}
\usepackage{ulem}
\usepackage[small,bf]{caption}
\usepackage{cite}
\usepackage[usenames,dvipsnames]{xcolor}
\usepackage{subfigure}
\usepackage{verbatim}

\allowdisplaybreaks

\numberwithin{equation}{section}

\newcounter{mysubequation}[equation]

\DeclareMathOperator{\diff}{\text{d}}

\begin{document}
\begin{titlepage}

\begin{center}
{
\bf\LARGE Detection Prospects of Dark Matter in\\[0.3em]
Einstein Telescope
}
\\[8mm]
Chuan-Ren~Chen\footnote[1]{crchen@ntnu.edu.tw} and Chrisna~Setyo~Nugroho\footnote[2]{setyo13nugros@ntnu.edu.tw}  
\\[1mm]
\end{center}
\vspace*{0.50cm}

\centerline{\it Department of Physics, National Taiwan Normal University, Taipei 116, Taiwan}
\vspace*{1.20cm}

\begin{abstract}
\noindent
We improve the calculations of the elastic motion induced by the dark matter hits on the surface of the mirror equipped with the interferometer for gravitational waves detection. We focus on the discovery potential of such a dark matter signal on the third-generation European gravitational waves detector, the Einstein Telescope. By taking the thickness of mirror into account, more than one resonances are predicted in the sensitive regime of high frequency interferometer. When mass of dark matter is heavier than a few PeV or is highly boosted, the signal-to-noise ratio could exceed one, and Einstein Telescope should be about to detect this dark matter signal.
\end{abstract}

\end{titlepage}
\setcounter{footnote}{0}

\section{Introduction}

The existence of dark matter (DM) provides the explanations for several astronomy observations, e.g. rotational curves of galaxies. However, except for gravitational effects, we know very little about its physical properties, including spin and mass. In other words, the DM particle could  be a boson or a fermion, and its mass could be as light as below eV or  could be heavier TeV. Even though no conclusive evidence of DM signal has been found in the laboratory, some observations, if interpreted as signals of DM, give us hits about the mass of DM. 
The recoil energy detection in XENON1T experiment~\cite{XENON:2020rca} may imply a dark matter with mass of $O(1)$ keV; the excesses of cosmic-ray positron observed by PAMELA~\cite{PAMELA:2008gwm} and AMS-02~\cite{AMS:2014bun} suggest that the mass of dark matter should be heavier than a few hundreds GeV; while the observations of PeV neutrino events by IceCube point to a super heavy dark matter. 

To claim the discovery of dark matter relies on the detection of recoil energy of either electron or nucleus inside the detector caused by the dark matter in the scattering, e.g. XENON1T experiment. 
However, interestingly, the technology used in the discovery of gravitational waves (GWs) by LIGO and VIRGO in 2016~\cite{Abbott:2016blz} provides another method for dark matter search. 
The idea for GWs detection is that as GWs pass through Earth, the tidal force causes tiny shifts in the length of LIGO/VIRGO arms, which can be detected by the laser interferometer. Similarly, when dark matter particles pass through the interferometer and hit the mirror of the interferometer, the recoil of the mirror may generate detectable vibration signals. 
The geometric properties of mirror and the mass of dark matter particle determine the characteristic frequencies. 
Details of such a scenario have been discussed in Ref.~\cite{Tsuchida:2019hhc}.
In this paper, we take into account the thickness of mirror and the noise caused by the thermal fluctuations, which have been neglected.   
We consider LIGO/VIRGO, KAGRA and the proposed Einstein Telescope (ET), and focus on ET since it's interferometer for high frequency is more optimistic to detect this signal, as will be shown later.
Actually, various approaches to search for DM with GW detectors have been discussed in the literature, see, for instance,~\cite{Lee:2020dcd,Seto:2004zu, Adams:2004pk, Riedel:2012ur, PhysRevLett.114.161301,
	Arvanitaki:2015iga, Stadnik:2015xbn, Branca:2016rez, Riedel:2016acj, Hall:2016usm,
	Jung:2017flg, Pierce:2018xmy, Morisaki:2018htj, Grote:2019uvn, Badurina:2019hst, 
	Graham:2017pmn, Coleman:2018ozp, Cheng:2019vwy, Bertoldi:2019tck, Dimopoulos:2008sv, 
	Graham:2012sy, Graham:2016plp, Canuel:2017rrp, Canuel:2019abg, Zhan:2019quq, Dimopoulos:2008hx, Kawasaki:2018xak, Monteiro:2020wcb}.

The rest of this paper is organized as follows. In Section~\ref{sec:DM Signal}, we give a detail derivation of the signal due to the collision between dark matter particles and mirror.  In Section~\ref{sec:noise}, we discuss the noises with emphasis on internal thermal noises. We show the signal to noise ratio in Section~\ref{sec:SNR}. Our summary and conclusions are presented in Section~\ref{sec:Summary}.

\section{Dark Matter Signal}
\label{sec:DM Signal}

We focus on the elastic oscillation of the mirror due to the hits of DM particles. We assume that
the test mass used in the interferometer (TM) collides with a DM
particle at $t=0$. During the collision, the DM  transfer a recoil momentum $q_{R}$ to the TM.
Following Refs.  \cite{Tsuchida:2019hhc,Lee:2020dcd}, the external force can be given by 
\begin{align}
\label{eq:FDM}
F_{DM} = q_{R}\,\delta(t).
\end{align} 
This collision induces transverse deflection $z_{\rm{Elas}}$ on the
surface of the mirror and causes the elastic oscillation of the
mirror. For $t>0$, the corresponding equations of motion of the TM with the mass $M_{T}$ due to this collision are~\cite{Mindlin:1954aip,Wen:2008wne}
\begin{align}
\label{eq:Mindlin}
\frac{\partial M_{r}}{\partial r} + \frac{1}{r}\,\frac{\partial M_{r\theta}}{\partial \theta} + \frac{M_{r}-M_{\theta}}{r} - Q_{r} - \frac{\rho h^{3}}{12} \frac{\partial^{2} \psi_{r}}{\partial t^{2}} &= 0 \nonumber 
\\
\frac{\partial M_{r\theta}}{\partial r} + \frac{1}{r}\,\frac{\partial M_{\theta}}{\partial \theta} + \frac{2 M_{r\theta}}{r} - Q_{\theta} - \frac{\rho h^{3}}{12} \frac{\partial^{2} \psi_{\theta}}{\partial t^{2}} &= 0 \nonumber
\\
\frac{\partial Q_{r}}{\partial r} + \frac{1}{r}\,\frac{\partial Q_{\theta}}{\partial \theta} + \frac{Q_{r}}{r} - \frac{\rho h  \omega_{e}}{Q_{M}} \frac{\partial z_{\rm{Elas}} } {\partial t} - \rho h \frac{\partial^{2} z_{\rm{Elas}}}{\partial t^{2}} &= 0 , \,
\end{align}
with the following initial conditions on $z_{\text{Elas}}$
\begin{align}
\label{eq:initz}
z_{\rm{Elas}} (r,\theta,t=0)=0 ,\,\,\, 
\dot{z}_{\text{Elas}}(r,\theta,t=0) =  \frac{q_{R}}{M_{T}}\, \pi\,a^{2} \delta(\vec{r} - \vec{r}_{0})\, ,
\end{align}
where $\dot{z}_{\text{Elas}} \equiv \frac{\partial z_{\text{Elas}}}{\partial t}$.
Here $h$, $a$, $\rho$, $Q_{M}$, and $\omega_{e}$ denote the thickness, the radius,
the mass density, the quality factor, and the angular eigenfrequency
of the mirror. The vector $\vec{r}_{0} = (r_{0},\theta_{0})$
specifies the location on the mirror where the DM particle hits. The components
of the mirror moments $M_{i}$ and shearing forces $Q_{i}$ can be
expressed in terms of the transverse deflection $z_{\rm{Elas}}$ as
well as the angular bending rotation of the normal to the
neutral surface in radial and circumferential directions, $\psi_{r}$, and $\psi_{\theta}$ as~\cite{Mindlin:1954aip}

\begin{align}
\label{eq:moments}
M_{r} &= D \left[\frac{\partial \psi_{r}}{\partial r} + \frac{\nu}{r}\left( \psi_{r} + \frac{\partial \psi_{\theta}}{\partial \theta} \right) \right] \, \nonumber
\\
M_{\theta} &= D \left[ \frac{1}{r}\left( \psi_{r} + \frac{\partial \psi_{\theta}}{\partial \theta} \right) + \nu \, \frac{\partial \psi_{r}}{\partial r} \right] \, \nonumber
\\
M_{r\theta} &= \frac{D}{2} (1 - \nu) \left[\frac{1}{r} \left( \frac{\partial \psi_{r}}{\partial \theta} - \psi_{\theta} \right) + \frac{\partial \psi_{\theta}}{\partial r} \right]  \, \nonumber
\\
Q_{r} &= \kappa^{2} G h \left( \psi_{r} + \frac{\partial z_{\rm{Elas}}}{\partial r} \right) \, \nonumber
\\
Q_{\theta} &= \kappa^{2} G h \left( \psi_{\theta} + \frac{1}{r}\frac{\partial z_{\rm{Elas}}}{\partial \theta} \right) \, ,
\end{align}
where the flexural rigidity of the mirror $D$ is written in terms of
the Young's modulus $E$ and Poisson's ratio $\nu$ as $D = Eh^{3}/
12(1 - \nu^{2})$. In Eq.~\eqref{eq:moments}, $G$ stands for the shear modulus and $\kappa^{2}= \pi^{2}/12$ is the shear coefficient. 
Note that the Eq.~\eqref{eq:Mindlin} is different from Kirchhoff thin plate theory~\cite{leissa:1969vib,chakraverty:2008vib}
used in Ref.~\cite{Tsuchida:2019hhc}.

The general solution of Eq.\eqref{eq:Mindlin} for $z_{\text{Elas}}$ can be written as
\begin{align}
\label{eq:solMindlin}
z_{\rm{Elas}} (r,\theta,r_{0},\theta_{0},t) &= \Theta(t)\,\frac{1}{\sqrt{1 - \frac{1}{4 Q^{2}_{M}}}} \nonumber \\
&\times \sum^{m,n=\infty}_{m,n=0}  K_{mn}(r_{0},\theta_{0}) \, W_{mn} (r,\theta)\, \text{exp} \left[-\frac{\omega_{mn}}{2Q_{M}}\, t \right] \text{sin}\left( \omega_{mn} \sqrt{1 - \frac{1}{4 Q^{2}_{M}}} t \right) 
\end{align}
where $m$ is the number of nodal diameters, $n$ denotes the number
of nodal circles, and $\omega_{mn}$ corresponds to the
eigenfrequencies for each mode. The coefficient $K_{mn}(r_{0},\theta_{0})$ depends on the
location of the DM hit on the mirror\footnote{This implies that the transverse deflection of the mirror $z_{\rm{Elas}}$ becomes $r_{0}$ and $\theta_{0}$ dependent as explicitly written in Eq.\eqref{eq:solMindlin}.}. The function $W_{mn}(r, \theta)$ can be
obtained by redefining the problem in terms of three potentials $w_{1}, w_{2}, w_{3}$ as~\cite{Irie:1980NaturalFO}

\begin{align}
\label{eq:redef}
\psi_{r} &= (\sigma_{1} - 1) \, \frac{\partial w_{1}}{\partial r} + (\sigma_{2} - 1) \, \frac{\partial w_{2}}{\partial r} + \frac{1}{r} \frac{\partial w_{3}}{\partial \theta} \nonumber
\\
\psi_{\theta} &= (\sigma_{1} - 1) \, \frac{1}{r}\frac{\partial w_{1}}{\partial \theta} + (\sigma_{2} - 1) \, \frac{1}{r}\frac{\partial w_{2}}{\partial \theta} -  \frac{\partial w_{3}}{\partial r} \nonumber
\\
W &= w_{1} + w_{2} , \,
\end{align}
where we have suppressed the  $m$ and $n$ indices. These potentials $w_{i}$ satisfy the following equations in polar coordinate

\begin{align}
\label{eq:wiEOM}
(\nabla^{2} + \delta^{2}_{i}) w_{i} = 0 \, , 
\end{align}
where $i$ runs from 1 to 3 and the Laplacian operator $\nabla^{2} = \partial^{2}/\partial \chi^{2} + (1/\chi) \, \partial/\partial \chi + (1/\chi)^{2} \, \partial^{2}/\partial \theta^{2}$ is expressed in term of
dimensionless quantity $\chi = r/a$. In Eq.\eqref{eq:redef} and Eq.\eqref{eq:wiEOM} we have introduced  dimensionless parameters
as

\begin{align}
\label{eq:dimparams}
\delta^{2}_{1}, \delta^{2}_{2} &= \frac{1}{2} \lambda^{2} \left\lbrace  R + S \pm \left[ \left( R - S \right)^{2} + 4 \lambda^{-2} \right]^{1/2} \right\rbrace \nonumber
\\ 
\delta^{2}_{3} &= \frac{2}{(1-\nu)} \left( R \lambda^{2} - S^{-1}\right) \nonumber
\\ 
\sigma_{1}, \sigma_{2} &= (\delta^{2}_{2}, \delta^{2}_{1}) (R \lambda^{2} - S^{-1})^{-1} \nonumber
\\ 
R &= (h/a)^{2} / 12  \nonumber
\\ 
S &= D/ (\kappa^{2} G a^{2} h) \, =\left[2/ (\pi^{2} (1 - \nu)) \right] (h/a)^{2} \,.
\end{align}
These parameters are the function of the dimensionless frequency parameter $\lambda$ (or more accurately $\lambda_{mn}$) as

\begin{align}
\label{eq:lambda}
\lambda_{mn} = \omega_{mn}\, a^{2} \left(\frac{\rho h}{D}\right)^{1/2} \, ,
\end{align}
where we restore the $m$ and $n$ indices.  
The general solutions of Eq.\eqref{eq:wiEOM} are given in terms of the product of the radial and angular function as
\begin{align}
\label{eq:wiSol}
w_{1}  &= A_{1} R_{m}\left( \Delta_{1}(\lambda_{mn}) \, \chi \right) \text{cos}(m\theta) \nonumber
\\
w_{2}  &= A_{2} R_{m}\left( \Delta_{2}(\lambda_{mn}) \, \chi \right) \text{cos}(m\theta) \nonumber
\\
w_{3}  &= A_{3} R_{m}\left( \Delta_{3}(\lambda_{mn}) \, \chi \right) \text{sin}(m\theta) \, ,
\end{align} 
where $\Delta_{i}$ is a function of $\lambda_{mn}$. Furthermore, it also depends on the sign of $\delta^{2}_{i}$ as~\cite{Xiang:2005xng}
\begin{subnumcases}{\Delta_{i}=}
  \delta_{i} \,\,\, \, \,  \text{if} \, \, \,\,\,  \delta^{2}_{i} \geq 0 \\
  \text{Im}(\delta_{i}) \,\,\,\,\, \text{if} \,\,\, \, \, \delta^{2}_{i} < 0\,.
\end{subnumcases}
The function $R_{m}$ is given in terms of the Bessel function of the first kind $J_{m}$ and its modified version $I_{m}$ as
\begin{subnumcases}{R_{m}\left( \Delta_{i} \, \chi \right)=}
  J_{m}(\Delta_{i} \, \chi)\,\,\,\,  \text{if} \, \, \,  \delta^{2}_{i} \geq 0 \\
  I_{m}(\Delta_{i} \, \chi)\,\,\,\,\,  \text{if} \, \, \, \delta^{2}_{i} < 0\, . 
\end{subnumcases}

The frequency
parameter $\lambda_{mn}$ is obtained by assuming that the edge
of the
mirror used in GW detectors are free.  This sets the boundary conditions at $r=a$ as
\begin{align}
\label{eq:BC}
M_{r}\rvert_{r=a} = M_{r\theta}\rvert_{r=a} =Q_{r}\rvert_{r=a} = 0 \, .
\end{align}
Substituting Eq.\eqref{eq:solMindlin}, Eq.\eqref{eq:redef} and Eq.\eqref{eq:moments} into the boundary conditions in Eq.\eqref{eq:BC}, we arrive at the following matrix equation
\begin{align}
\label{eq:Meigen}
\begin{pmatrix}
C_{11} & C_{12}
& C_{13}  \\
C_{21}
& 
C_{22} 
&  C_{23} \\
C_{31}  & C_{32} &  C_{33}  
\end{pmatrix} \, \begin{pmatrix}
A_{1} \\
A_{2} \\
A_{3}
\end{pmatrix} = \begin{pmatrix}
0 \\
0 \\
0
\end{pmatrix} \, .
\end{align}
This can be written in vector notation $\textbf{C} \vec{A} = \vec{0}$. The elements of the $\textbf{C}$ matrix are
\begin{align}
C_{1j} &= (\sigma_{j} - 1) \left[ R^{''}_{m}(\Delta_{j}) + \nu \, R^{'}_{m}(\Delta_{j}) - \nu \, m^{2} R_{m}(\Delta_{j}) \right] \, , \nonumber
\\
C_{13} &= m \, (1 - \nu) \left[ R^{'}_{m}(\Delta_{3}) - R_{m}(\Delta_{3}) \right] \nonumber \, ,
\\
C_{2j} &= -2\, m \,(\sigma_{j} - 1) \left[ R^{'}_{m}(\Delta_{j}) - R_{m}(\Delta_{j}) \right] \nonumber \, ,
\\
C_{23} &= -\left[ R^{''}_{m}(\Delta_{3}) - R^{'}_{m}(\Delta_{3}) + m^{2} \, R_{m}(\Delta_{3})\right] \nonumber \, ,
\\
C_{3j} &= \sigma_{j} R^{'}_{m}(\Delta_{j}) \nonumber \, ,
\\
C_{33} &= m\, R_{m}(\Delta_{3}) \, ,
\end{align} 
where $R^{'}_{m}(\Delta_{i}\chi)=\frac{\partial R_{m}(\Delta_{i}\chi)}{\partial\chi}$ and the index $j$ runs from 1 to 2. The non-trivial solution of
Eq.\eqref{eq:Meigen} requires the determinant of the matrix $\textbf{C}$ to vanish, which further sets the values of eigenfrequency.

For the existing  as well as the proposed third generation GW experiments, the detailed properties of their mirrors
are summarized in table \ref{tab:GWMirror}. 
 For LIGO, VIRGO, KAGRA,
and low frequency Einstein Telescope (ET LF) the lowest 
eigenfrequencies are 10.74, 10.74, 15.33, and 15.75 kHz, respectively.
 These frequencies lie outside their detection band.
This can be understood from Eq.~\eqref{eq:lambda} and table \ref{tab:GWMirror} where the value of the eigenfrequency is
proportional to the ratio of thickness to  radius $h/a$,
the Young's modulus $E$, and the mass density of the mirror $\rho$. 
On the other hand, the eigenfrequencies of the mirror equipped with the high frequency
Einstein Telescope (ET HF)  are located inside the detection band as tabulated in table \ref{tab:wmn}. We find that there are four resonance peaks inside the ET HF detection band. 
This is different from the study of Ref.~\cite{Tsuchida:2019hhc} which found one resonance peak at 7.24 kHz. 
Furthermore, the thin mirror proposed in Ref.~\cite{Tsuchida:2019hhc} could generate more resonance peaks inside ET detection band.
However, as will be shown below, this is not a suitable option since the resonance peaks are also excited by internal thermal noise of the mirror rendering the detector to lose its sensitivity. From here on, we
focus on studying the elastic oscillations of the mirror in ET HF induced by DM hit.   

\begin{table}[h!]
\centering
\begin{tabular}{|c c c c c|} 
 \hline
  & LIGO, VIRGO & KAGRA & ET LF & ET HF \\ [0.5ex] 
 \hline \hline
 Material & Fused Silica & Sapphire & Silicon & Fused Silica \\ 
 Mirror Mass, $M_{T}$ [kg] & 40 & 22.8 & 211 & 200\\
 Thickness, $h$ [cm] & 20 & 15 & 50 & 30\\
 Radius, $a$ [cm] & 17.5 & 11 & 22.5 & 31 \\ 
 Mass Density, $\rho$ [g/$\text{cm}^{3}$] & 2.20 & 4.00 & 2.33 & 2.20 \\
 Young's Modulus, $E$ [GPa] & 72 & 400 & 188 & 72 \\
 Poisson's Ratio, $\nu$  & 0.17 & 0.30 & 0.22 & 0.17 \\[1ex]
 \hline
\end{tabular}
\caption{The properties of the mirror used in the interferometers for different  GW experiments \cite{Tsuchida:2019hhc,ET Science Team: 2010ets}.}
\label{tab:GWMirror}
\end{table}

\begin{table}[h!]
\centering
\begin{tabular}{|c c c c c|} 
 \hline
  & $m = 0$ & $m = 1$ & $m = 2$ & $m = 3$ \\ [0.5ex] 
 \hline \hline
 $n = 0$ & ... & ... & 0.240 & 0.485 \\ 
 $n = 1$ & 0.514 & 0.601 & 0.791 & 0.897\\
 $n = 2$ & 0.548 & 0.635 & 1.100 & 1.338\\
 $n = 3$ & 0.643 & 0.820 & 1.271 & 1.455 \\[1ex] 
 \hline
\end{tabular}
\caption{The numerical value of eigenfrequencies ($\frac{\omega_{mn}}{2\pi}$) of the ET HF for the first few of each $m$ and $n$ in unit of $\times 10^{4}$ [Hz]. The quality factor for HF ET mirror is $2.5 \times 10^{9}$~\cite{ET Science Team: 2010ets}.}
\label{tab:wmn}
\end{table}

The function $W_{mn}(r, \theta)$ can be determined by expressing the
coefficient $A_{2}$ and $A_{3}$ in terms of $A_{1}$ via Eq.\eqref{eq:Meigen} and using Eq.\eqref{eq:redef} as

\begin{align}
\label{eq:WmnSol}
W_{mn}(r, \theta) = A_{1} \left[R_{m}\left( \frac{\Delta_{1}(\lambda_{mn})}{a} r  \right) - \left[ \frac{C_{31}C_{23} - C_{21} C_{33}}{C_{32}C_{23} - C_{22} C_{33}} \right] R_{m}\left( \frac{\Delta_{2}(\lambda_{mn})}{a} r  \right) \right] \text{cos}(m \theta) \, .
\end{align}
The transverse deflection of the mirror $z_{\rm{Elas}} (r,\theta,t)$ 
can be determined by inserting Eq.\eqref{eq:WmnSol} back into Eq.\eqref{eq:solMindlin}, with the redefinition of $ A_{1}\,K_{mn} \rightarrow K_{mn}$. In GW experiments, the signal of the gravitational waves as well as the noises present in the
interferometer  are analyzed in terms of strain amplitude in
frequency domain. Therefore, one needs to do the Fourier transform
\begin{align}
\tilde{z}_{\text{Elas}}(r,\theta,r_{0},\theta_{0},f)=\int^{\infty}_{-\infty} z_{\rm{Elas}} (r,\theta,r_{0},\theta_{0},t) \, e^{-\text{i}\,2\pi f t}\, dt \, ,
\label{eq:fourier}
\end{align}
of the DM signal and evaluate its absolute value in order to write the corresponding strain amplitude of the signal
\begin{align}
\label{eq:zf}
|\tilde{z}_{\text{Elas}}(r,\theta,r_{0},\theta_{0},f)| = \frac{1}{2\pi} \sum^{m=\infty}_{m=0} \, \sum^{n=\infty}_{n=0}K_{mn}(r_{0},\theta_{0})\, W_{mn}(r,\theta) \, \frac{f_{mn}}{\sqrt{(-f^{2} + f^{2}_{mn})^{2} + (\frac{f\,f_{mn}}{Q_{M}})^{2}}} \, ,
\end{align}
where $f_{mn} \equiv \frac{\omega_{mn}}{2\pi}$ is the eigenfrequency
of the mirror expressed in hertz. We see that at resonance frequency
$f = f_{mn}$, the elastic oscillation induced by DM hit has
pronounced peaks.

To determine the coefficient $K_{mn}$, we employ the momentum
conservation~\cite{Tsuchida:2019hhc}
\begin{align}
\label{eq:momConser}
q_{R}\, \delta(\vec{r} - \vec{r}_{0} ) =  2\pi \rho h \sum^{m=\infty}_{m=0} \sum^{n=\infty}_{n=0} K_{mn} W_{mn}(r,\theta) f_{mn} \, ,
\end{align}
where $\vec{r}_{0}$ specifies the location of the DM hit on the mirror.
We multiply to both side of Eq.\eqref{eq:momConser} by $W_{kl}(r,\theta)$ and further integrate over the mirror surface 
\begin{align}
\label{eq:Kmn}
q_{R}\, W_{kl}(r_{0},\theta_{0}) &= 2\pi \rho h K_{kl} f_{kl}\int^{2\pi}_{0} \int^{a}_{0} W^{2}_{kl}(r,\theta) r\,dr \, d\theta \,, \nonumber
\\
K_{mn}(r_{0},\theta_{0}) &= \frac{q_{R}\, W_{mn}(r_{0},\theta_{0})}{2\pi \rho h f_{mn} \, \int^{2\pi}_{0} \int^{a}_{0} W^{2}_{mn}(r,\theta) r\, dr \,  d\theta} \, ,
\end{align}
where we change the index $k, l$ back to $m, n$ in the last line.
The laser beam used in a typical GW detector is focused at the center of the
mirror to monitor the differential change of the interferometer arm.  Therefore, we evaluate the elastic oscillation at the center of the
mirror which corresponds to $m = 0$ mode or the axis-symmetric mode. In this case, the coefficient $K_{mn}$ reads
\begin{align}
\label{eq:K0n}
K_{0n}(r_{0},\theta_{0}) = \frac{q_{R} W_{0n}(r_{0},\theta_{0})}{4\pi^{2} \rho h f_{0n} \int^{a}_{0} W^{2}_{0n}(r) \, r\, dr} \,.
\end{align}

The magnitudes of $|\tilde{z}_{\text{Elas}}|$ at resonance
frequencies $f = f_{0n} $ evaluated at $r = 0$ for different nodal
circle modes $n$ and various collision points on the mirror $r_{0}$
are collected in table \ref{tab:Ztildemn}. 
\begin{table}[h!]
\centering
\begin{tabular}{|c |c| c c c c c c|} 
 \hline
  & $f_{0n}$ [kHz] &$r_{0} = 0.0\,a$ & $r_{0} = 0.1\,a$ & $r_{0} = 0.2\,a$ & $r_{0} = 0.3\,a$ & $r_{0} = 0.4\,a$ & $r_{0} = 0.5\,a$ \\ [0.5ex] 
 \hline \hline
 $n = 1$ & 5.136 & 282.8 & 265.1 & 216.8 & 150.9 & 84.67 & 33.03 \\ 
 $n = 2$ & 5.482 & 253.6 & 236.2 & 189.4 & 127.0 & 66.41 & 22.21\\
 $n = 3$ & 6.431 & 192.1 & 175.4 & 131.9 & 77.62 & 31.08 & 4.828\\
 $n = 4$ & 9.072 & 163.3 & 139.8 & 84.52 & 30.16 & 2.431 & 3.068 \\[1ex] 
 \hline
\end{tabular}
\caption{The values of $|\tilde{z}_{\text{Elas}}(r = 0,\theta=0,r_{0},\theta_{0}, f = f_{0n})|\times 10^{-23}$ at different collision points $r_{0}$ for $q_{R} \approx m_{\text{DM}} \, v = 1 \,\text{GeV}/\text{c}^{2} \times 220 \, \text{km/s}$ in unit of $\text{cm}\,\text{Hz}^{-1}$. The resonance frequency for each mode is shown in the second column.}
\label{tab:Ztildemn}
\end{table}
As can be seen from this table, for each $n$ mode, the
highest transverse displacement is reached when DM hits the center of the mirror
($r_{0} = 0$). 
To compare the DM signal with ET sensitivity, we need to calculate the strain amplitude induced by DM hit. It is given by
\begin{align}
\label{eq:DMStrain}
\tilde{h}_{\text{DM}}(r,\theta,r_{0},\theta_{0},f) = \sqrt{\frac{4\,f}{L^{2}}}\, |\tilde{z}_{\text{Elas}}(r,\theta,r_{0},\theta_{0}, f)| \,,
\end{align} 
where $L$ denotes the arm length of the interferometer. We aware
that the resonance modes discussed here can also be excited by the noise.
We discuss the relevant noise that may overcome the DM signal in the next section. 

\section{Relevant Noise}
\label{sec:noise}

In this section we briefly discuss the noise that affects the 
sensitivity of ET. The details of the noise components can be found
in~\cite{ET Science Team: 2010ets} and references therein, therefore, we will not repeat it here. We only highlight the
dominant noise component that would potentially suppress the DM signal
under consideration. 

Einstein telescope employs a new strategy called xylophone technique
in its design. It combines two interferometers operating in
different frequency regimes. One interferometer (ET LF) is cooled down to the
cryogenic temperature at $10$ K with low laser power to reduce radiation
pressure noise. This is designed to have a high sensitivity at low frequency regime.
On the other hand, the second interferometer (ET HF) 
works at room temperature ($290$ K), and a high-power laser is utilized to reduce the shot noise. 
 This is done to reach a high
sensitivity at high frequencies. As a result, this technique allows
ET to gain good sensitivities in both low and high frequencies 

\begin{figure}
	\centering
	\includegraphics[width=0.8\textwidth]{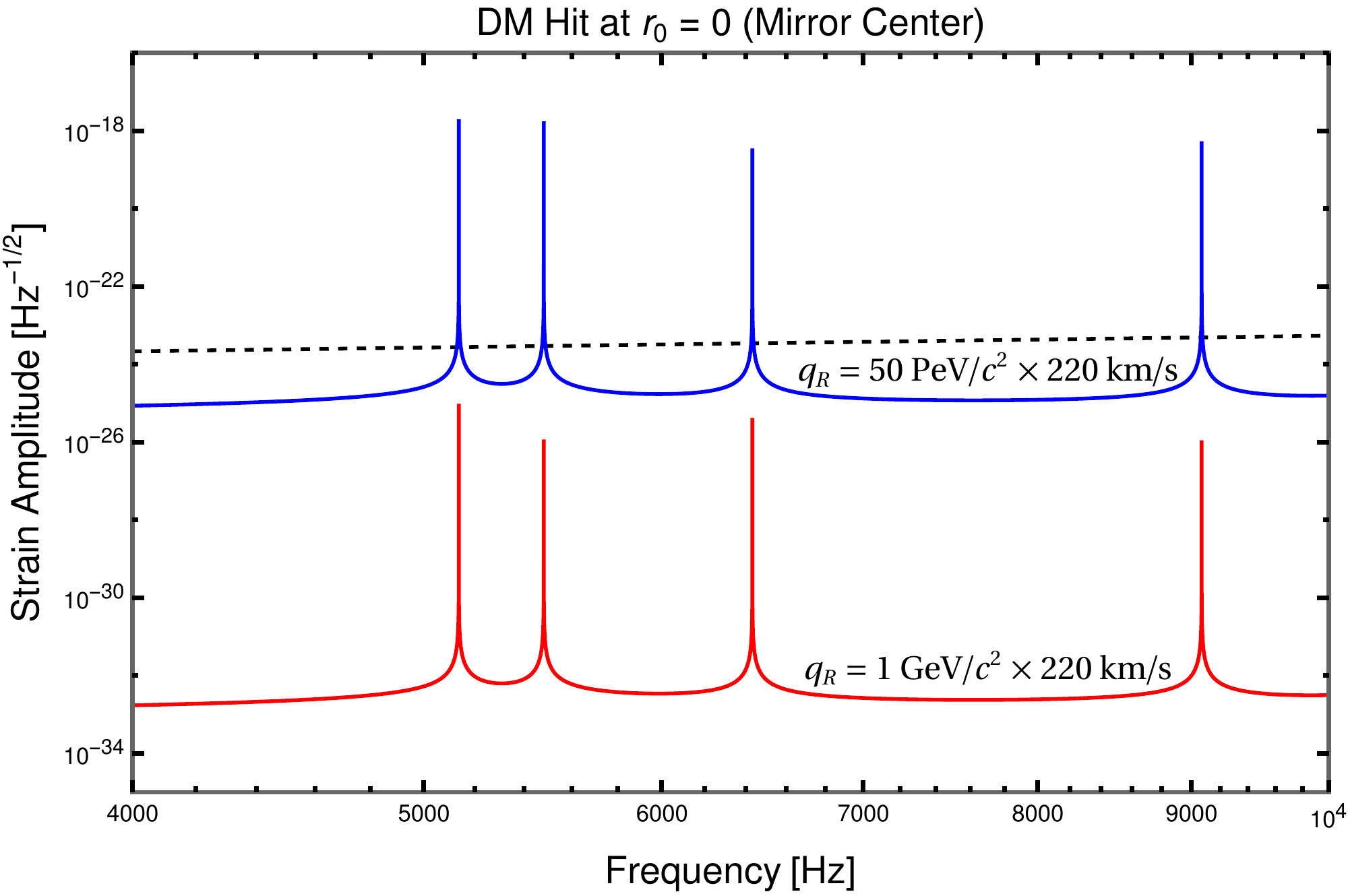}
	\caption{The strain sensitivity of ET~\cite{Ligo:2020doc} (black dashed line) and the hypothetical DM signals for light DM with mass of 1 GeV/$\text{c}^{2}$ (red line) as well as heavy DM with mass of 50 PeV/$\text{c}^{2}$ (blue line). 
The velocity of dark matter particle is taken to be $220 \text{ km/s}$.}
	\label{fig:DMMindlin}
\end{figure}

At low frequencies below $10$ Hz, the dominant noise components are seismic noise and
gravity-gradient noise. 
In the intermediate regime up to $500$ Hz, the relevant noises are suspension thermal noise and internal thermal noise (mirror thermal noise).
The internal thermal noise excites the eigenfrequencies of the mirror and may potentially hide the DM signal. In fact, the internal thermal noise of
the mirror with the power spectral density (PSD) given by~\cite{Saulson:1990jc} 
\begin{align}
\label{eq:ThPSD}
x^{2}_{\text{th}}(f) = \frac{8\,k_{B}\,T}{(2\pi)^{3}} \,\sum^{m,n=\infty}_{m,n=0} \frac{\phi_{mn}(f) \, f^{2}_{mn}}{M_{T}\, f\, \left[ (f^{2} - f^{2}_{mn})^{2} + \phi^{2}_{mn}(f)\, f^{4}_{mn} \right] } \, 
\end{align}
mimicks the DM signal at $f = f_{mn}$. Here, $k_{B}, T$, and $M_{T}$
stand for the Boltzman constant, the temperature, and the mirror
mass, respectively. In many materials, the mechanical loss angle $\phi_{mn}$ is approximately a constant over a wide band of
frequencies. It characterizes the internal damping of the mirror and is
inversely proportional to the quality factor of the mirror, namely $\phi_{mn} \propto 1/Q_{M}$. 
Note that as temperature drops down to zero or the loss
angle vanishes, there would be no internal thermal noise, as it should be. The corresponding strain amplitude due to this noise is 
\begin{align}
\label{eq:hTh}
\tilde{h}_{\text{th}}(f) = \sqrt{\frac{4\,x^{2}_{\text{th}}(f)}{L^{2}}} \, .
\end{align}
On the other hand, the most dominant noise component in high frequency regime
above 500 Hz comes from the photon shot originated from the
phase fluctuation of the laser beam. This noise affects the readout of
interference fringes at dark port of the interferometer where the GWs
signal is detected. Several techniques can be applied to reduce the
shot noise. ET HF plans to use the squeezed states of the photons to
optimize the sensitivity in certain frequency regime below the standard quantum limit.
As an additional remark, ET employs ultra high vacuum (UHV) to reduce the surrounding gas pressure at $10^{-10} $ mbar [38]. The corresponding quality factor should be much higher than the mirror quality factor considered here. Therefore, the viscous damping due to the ambient gas collisions is insignificant~\cite{Saulson:1990jc}.

Both internal thermal noise and photon shot noise are the dominant
noises that potentially conceal the presence of our DM signals. In
Fig.\ref{fig:DMMindlin} we show the strain amplitude curve of ET
that includes all the noise components~\cite{Ligo:2020doc}.
The DM signals for two benchmark masses of $1\, \text{GeV}/\text{c}^{2}$ and
$50\, \text{PeV}/\text{c}^{2}$, representing light and heavy DM, respectively, are also
shown. The height of the resonance peaks seems to be different in
both cases. This is due to the difficulty to plot very narrow peaks
around the resonances. As one can see, the dark matter signal of
heavy DM lies
above the ET sensitivity curve at resonant frequencies while the
light DM signal is buried under the noises.

\section{Signal to Noise Ratio}
In this section, we would like to analyze the detectability of the
DM signal discussed previously in term of signal-to-noise ratio
(SNR), $\varrho^2$. 
Basically, this is done by integrating the ratio of the DM
signal over noise in the corresponding frequency domain.
Furthermore, one applies the appropriate filter to improve the SNR which is optimally given by\cite{Moore:2014lga}  

\begin{equation}
\varrho^2 \equiv  \int_{f_\text{min}}^{f_\text{max}} \frac{\diff f}{f} \frac{S_{s}(f)}{S_{n}(f)} \;,
\label{eq:defSNR}
\end{equation} 
where $S_{s}(f) = \tilde{h}^{2}_{s}(f)$ and $S_{n}(f) = \tilde{h}^{2}_{n}(f)$ are the
PSD of the signal and the noise,
respectively. Here, the integral runs over the frequency range probed by
the
corresponding detector. In our case, the SNR reads
\begin{align}
\label{eq:opSNR}
\varrho^2(r,\theta, r_{0},\theta_{0}) = \int_{f_\text{min}}^{f_\text{max}} \frac{\diff f}{f} \frac{ \tilde{h}^{2}_{\text{DM}}(r,\theta,r_{0},\theta_{0}, f)}{\tilde{h}^{2}_n(f)} \;.
\end{align}
In the following, we take the axis-symmetric mode ($m = 0$) to evaluate the SNR in Eq.~\eqref{eq:opSNR}
as well as $r_{0} = \theta_{0} = 0$ which correponds to DM hit at
the center of mirror. Furthermore, since the laser is focused at the
center of the mirror, we set $r = 0$. This setup corresponds to the maximal $|\tilde{z}_{\text{Elas}}(r,\theta,r_{0},\theta_{0}, f)|$ as
can be
seen from table \ref{tab:Ztildemn}.
\label{sec:SNR}
\begin{figure}
	\centering
	\includegraphics[width=0.8\textwidth]{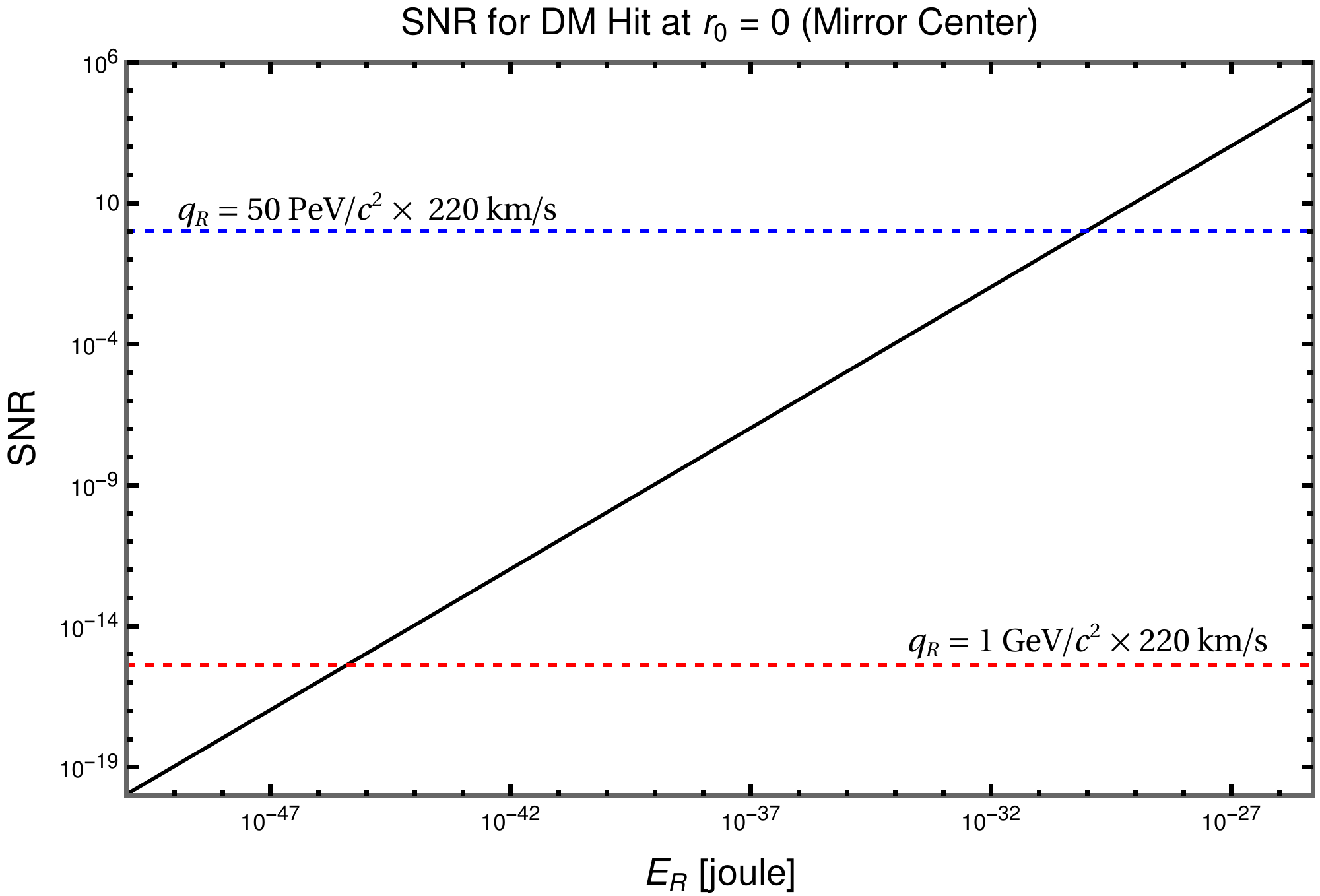}
	\caption{The SNR as a function of the recoil energy $E_{R}$ (solid black line) for Einstein Telescope. The corresponding SNR for the light DM (red dashed line, $q_{R}=1\, \text{GeV}/\text{c}^{2}\times 220\, \text{km/s}$) and heavy DM (blue dashed line, $q_{R}=50\, \text{PeV}/\text{c}^{2}\times 220\, \text{km/s}$) are  $4.20 \times 10^{-16}$ and $1.05$, respectively.}	
	\label{fig:SNRET}
\end{figure}
To get a better understanding about
the quantities that optimize the SNR, we make a simplification first
before evaluating the actual SNR. Let us consider under which
circumstances the DM signal overcomes the internal thermal noise.
Based on Fig.~\ref{fig:DMMindlin}, it is instructive to evaluate the
SNR around the DM signal peak. We see from Eq.~\eqref{eq:ThPSD} that
the PSD has a Lorentzian form peaked at $f=f_{0n}$ (for $m=0$) with the associated full-width at half maximum (FWHM) given by
\begin{align}
\label{eq:FWHMSimp}
f^{2}_{\text{max}} &= f^{2}_{0n}(1 + \phi_{0n})\;, \nonumber \\
f^{2}_{\text{min}} &= f^{2}_{0n}(1 - \phi_{0n})\;.
\end{align}
After expanding the $\tilde{h}_{\text{DM}}(f)$ and $\tilde{h}_{\text{th}}(f)$ around the resonance ($f \approx f_{0n}$) and setting $\phi_{0n} \approx \frac{1}{Q_{M}}$, the SNR becomes
\begin{align}
\label{eq:SNRTh}
\varrho_{\text{th}}^2 &= \int_{f_\text{min}}^{f_\text{max}} \frac{\diff f}{f} \frac{\tilde{h}^{2}_{\text{DM}}(0,0,0,0, f)}{\tilde{h}^{2}_n(f)} \approx \int_{f_\text{min}}^{f_\text{max}} \frac{\diff f}{f} \frac{\tilde{h}^{2}_{\text{DM}}(0,0,0,0, f)}{\tilde{h}^{2}_{\text{th}}(f)} =\int_{f_\text{min}}^{f_\text{max}} \diff f \, \frac{|\tilde{z}_{\text{Elas}}(0,0,0,0,f)|^{2}}{x^{2}_{\text{th}}(f)} \nonumber \\
&= \frac{\pi^{2}\,a^{4}}{(2\pi)^{3}}\,\frac{q^{2}_{R}}{2M_{T}}\,\frac{W^{4}_{0n}(0,0)}{\left(\int_{0}^{a}  W^{2}_{0n}(r) \,r\, \diff r\right)^{2}} \, \frac{1}{k_{B}\,T} \, \int_{f_\text{min}}^{f_\text{max}} \diff f \frac{f}{f^{2}_{0n}\,\phi_{0n}} \nonumber \\
&= \frac{a^{4}}{16\pi}\, \frac{E_{R}}{E_{\text{th}}} \, \left[ \frac{W^{2}_{0n}(0,0)}{\int_{0}^{a}  W^{2}_{0n}(r) \,r\, \diff r}\right]^{2}\, \frac{f^{2}_{\text{max}}-f^{2}_{\text{min}}}{2\,f^{2}_{0n}\,\phi_{0n}} =\frac{a^{4}}{\pi}\, \frac{E_{R}}{E_{\text{th}}} \, \left[ \frac{W^{2}_{0n}(0,0)}{4\,\int_{0}^{a}  W^{2}_{0n}(r) \,r\, \diff r}\right]^{2} \nonumber \\
&\approx \frac{1}{\pi} \, \frac{E_{R}}{E_{\text{th}}} \, \left[ \frac{J^{2}_{0}(0)}{2\left(J^{2}_{0}(\Delta_{1}(\lambda_{0n}))+J^{2}_{1}(\Delta_{1}(\lambda_{0n}))\right)}\right]^{2}\, ,
\end{align}  
where we have defined the recoil energy $E_{R} \equiv q^{2}_{R}/(2M_{T})$ and the thermal energy $E_{\text{th}} \equiv k_{B}T/2$ for single
degree of freedom. To obtain the last line, we set $\delta^{2}_{i} > 0$ and use the fact that the first term on the right hand side of Eq.~\eqref{eq:WmnSol} dominates over the second term. The integration boundary is given by Eq.~\eqref{eq:FWHMSimp}. 

Apart from the normal mode function, the SNR is proportional to the
ratio between dark matter recoil energy ($E_{R}$) and thermal energy ($E_{\text{th}}$). Thus, to optimize the SNR in this case, one needs a
lighter mirror and lower temperature. This is in qualitative
agreement with the toy model considered in~\cite{Lee:2020dcd}. However, lowering the mass of mirror 
will not be suitable for current GWs detector as it
enhances the radiation pressure noise at low frequency and further
limits the sensitivity of the detector. Furthermore, reducing the
temperature would significantly increase the loss angle $\phi_{mn}$
relevant for the thermal mirror noise~\cite{ET Science Team: 2010ets}. The only option left is to have DM with higher recoil momentum $q_{R}$.  

We see from the second line of Eq.~\eqref{eq:SNRTh} that the SNR is
proportional to $q^{2}_{R} = 2 M_{T} E_{R}$. The recoil energy can
be used to parameterize the SNR in any GWs detector. In ET HF, we
focus on the middle peak of Fig.~\ref{fig:DMMindlin} around 6431 Hz
to compute the SNR. For the integration boundary, one can use the
FWHM as before. However this is not recommended since the frequency
resolution of ET HF is bigger than the FWHM. Thus, we choose the following integration boundary
\begin{align}
\label{eq:fBound}
f_{\text{min}} = 6426 \,\text{Hz} \,\, \text{and} \,\, f_{\text{max}} = 6436 \,\text{Hz}.
\end{align}
The resulting SNR as a function of $E_{R}$ is shown in Fig.~\ref{fig:SNRET}. It is clear that, in order to achieve a considerable signal for detection, namely $\text{SNR} >1$, one needs the recoil energy higher than about $10^{-30}$ J. 
Also note that in our numerical results, we take the typical velocity about $220~\text{km/s}$ of dark matter
particles as the Earth passes through the dark matter halo. In this case, ET HF should be able to detect the signal of DM with mass heavier than  $\text{PeV}/\text{c}^{2}$, such as the super-heavy DM suggested in ~\cite{Hall:2016usm}. 
On the other hand,
the kinetic energy of light dark matter boosted by unknown sources
may reach the PeV regime as well\cite{Ema:2018bih, Cappiello:2019qsw, Guo:2020drq, Xia:2021vbz}. As a result, the recoil energy $E_R$
will be enhanced and makes the detectability of such dark matter particles in ET become feasible.

\section{Summary and Conclusions}
\label{sec:Summary}

We describe the scattering between DM particles and GWs detectors with the focus on the High Frequency Einstein Telescope (ET HF). We
show that DM induces elastic oscillation on the mirror employed in
the interferometer arm of ET. This motion excites the resonance frequencies of the mirror and can be detected in the
optomechanical setup used in ET.  

Our work improves the previous calculations of the DM signal in
two ways. First, by taking into account the thickness of the mirror
we found four resonance frequencies of the mirror inside ET HF
detection band as opposed to only a single resonance frequency found in the
previous study. Furthermore, we demonstrate that the same resonance
frequencies are excited by mirror thermal noise. This may conceal
the hypotetical DM signal and hence can not be neglected. To
overcome the thermal noise one needs to reduce both of the mirror
mass and its temperature. 

We consider the detection of light DM and heavy DM  with the mass of 1 $\text{GeV}/\text{c}^{2}$ and 50 $\text{PeV}/\text{c}^{2}$, respectively, as our benchmarks. We
find the linear relationship between the recoil energy induced by DM hit and the SNR.
The signal from heavy DM with the mass of the order of few PeV  could reach SNR $\gtrsim 1$ while the light DM would be buried by the thermal noise unless it is highly boosted. 


\section*{Acknowledgment}
\label{sec:Acknowledgment}
We would like to acknowledge the support of National Center for Theoretical Sciences (NCTS). This work was supported in part by the Ministry of Science and Technology (MOST) of Taiwan under Grant No.MOST 109-2112-M-003-004-, 110-2112-M-003-003- and 110-2811-M-003-505-.

\end{document}